\documentclass[11pt,prd,superscriptaddress,nofootinbib]{revtex4}
\usepackage[usenames, dvipsnames]{xcolor}
\usepackage
{amssymb, amsmath, amsfonts}
\usepackage{graphicx}

\begin{document}
		
	\titlepage
	
	\vspace{25mm}
	
	\begin{center}
		{\Large \bf Deuteron electromagnetic form factors and tensor polarization observables in the framework of the hard-wall AdS/QCD model}
		
		\vskip 1. cm
		{Narmin Huseynova $^{a}$\footnote{e-mail : nerminh236@gmail.com},
			Shahin Mamedov $^{a,b}$\footnote{e-mail : sh.mamedov62@gmail.com} and Jannat Samadov $^{c}$\footnote{e-mail: jannat.samadov@gmail.com}}
		
		\vskip 0.5cm
		{\it $^a\,$Institute for Physical Problems, Baku State University,
			Z.Khalilov 23, Baku, AZ-1148, Azerbaijan}\\
		{\it $^b\,$Institute of Physics, Azerbaijan National Academy of Sciences,
			H.Javid 33, Baku, AZ 1143, Azerbaijan}\\
		{\it $^c\,$ Shamakhy Astrophysical Observatory, Azerbaijan National Academy of Sciences, H.Javid 115, Baku, AZ 1000, Azerbaijan}\\
	\end{center}

	\centerline{\bf Abstract} \vskip 4mm
	
	We study the electromagnetic form factors and tensor polarization observables of the deuteron in the framework of the hard-wall AdS/QCD model. We find a profile function for the bulk twist $\tau=6$ vector field, which describes the deuteron on the boundary and fix the infrared boundary cut-off of AdS space in accordance with the ground state mass of the deuteron. We obtain the deuteron charge monopole, quadrupole, and magnetic dipole form factors and tensor polarization observables from the bulk Lagrangians for the deuteron and photon field interactions. We plot the momentum transfer dependence of the form factors and tensor polarization observables and compare our numerical results with those in the soft-wall model and experimental data.
	
	\vspace{1cm}
	\section{Introduction}
	
	The deuteron is formed mainly by ionizing deuterium with the emission of a photon: $d+\gamma\rightarrow d$. It also forms as a result of the capture of a neutron by a free proton in hydrogen contained matter, accompanied by the emission of a photon: $p+n\rightarrow d+\gamma$. The interaction vertex of these processes is described by electromagnetic (EM) form factors, which include the charge monopole $G_C(Q^2)$, quadrupole $G_Q(Q^2)$, and magnetic dipole $G_M(Q^2)$ form factors of the deuteron. We study the EM form factors and tensor polarization observables of the deuteron through the $e+d \rightarrow e+d$ electron-deuteron elastic scattering process.
	
	The EM form factors \cite{1,2,3,4,5,6,7,8,9,10,11,12,13,14,15,16,17,18,19,20,21,22,23,24,25,26,27,28} and tensor polarization observables \cite{3,17,19,21,24,29,30,31,32} of the deuteron have been studied experimentally \cite{33,34} and in the framework of several theoretical approaches. For instance, there are different relativistic models that describe deuteron EM form factors, which are based on phenomenological potential models, such as Nijmegen, JISP-16, CD-Bonn, Paris, Moscow, and Argonne \cite{11,12,15,16,28}. Deuteron EM form factors and tensor polarization observables have been calculated within the Born approximation approach (see Refs. \cite{1, 14, 17}) and studied theoretically  within  the  effective  Lagrangian \cite{4, 5}
	and  phenomenological Lagrangian approaches \cite{3, 7, 9}. In Ref. \cite{8}, deuteron EM form factors were investigated within a modified version of Weinberg's chiral effective field theory approach applied to a two-nucleon system. In Ref. \cite{10}, the EM form factors and low-energy observables of the deuteron were studied within the lightfront approach, where the deuteron is regarded as a weakly bound state of a proton and a neutron. In Ref. \cite{18}, the asymptotics of the EM form factors of the deuteron in the nuclear model and JLAB experiment were studied, where the deuteron was considered a two-nucleon system. After the introduction of holographic quantum chromodynamics (QCD), the EM form factors and tensor polarization observables of the deuteron were studied within the soft-wall model of anti-de Sitter (AdS)/QCD in Refs. \cite{19,20}. AdS/QCD models have been successfully applied to the study of the different couplings and form factors of hadrons \cite{19,20,35,36,37,38,39,40,41,42,43,44,45,46,47}. The holographic duality principle connects gauge theory in $d$-dimensional space with gravitational theory in $(d + 1)$-dimensional space and has great importance in solving many problems of QCD. Unlike quantum field theory, AdS/QCD models are used without restrictions on the momentum transfer when solving problems involving strong interactions. For this reason, AdS/QCD models have been successfully used in the study of strongly interacting quark-gluon plasma, as well as in the calculation of the strong interaction constants and form factors of hadrons. 

    Additionally, over the last decade, the soft-wall and hard-wall AdS/QCD models have improved through the consideration of appropriate Lagrangian terms or gravity background \cite{48,49,50,51,52,53,54,55,56,57,58,59,60,61,62,63,64,65}, for example, by considering the infrared (IR)-modified five-dimensional (5D) conformal mass and Yukawa coupling of the bulk baryon field with the bulk meson field, which has been shown to provide a consistent prediction for the mass spectra of mesons. A soft-wall AdS/QCD model with a modified 5D metric at the IR region was constructed to obtain a nontrivial dilaton solution, which incorporates chiral symmetry breaking and linear confinement. By taking the pion mass and decay constant as two input mass scales, the resulting predictions for the resonance states of mesons concurred remarkably with the experimentally confirmed resonance states; hence this model can lead to a consistent prediction for the mass spectra of resonance states in pseudoscalar, scalar, vector and axial-vector mesons \cite{51,52,53,64}. To implement confinement, the sign of dilaton in the IR soft-wall AdS/QCD model is changed, where the deformed model exhibits interesting properties, especially in describing chiral symmetry breaking \cite{48}. In Ref. \cite{49}, the predictive soft-wall AdS/QCD model with a modified 5D metric in the IR region was constructed to obtain a non-trivial dilaton solution for three flavor quarks $u$, $d$ and $s$, incorporating both chiral symmetry breaking and linear confinement. The thermal mass spectra of scalar and pseudoscalar mesons have also been studied within the IR-improved soft-wall AdS/QCD model in Ref. \cite{50}. 
	
	The principal idea of AdS/QCD is to set up models based on AdS/conformal field theory (CFT) that include key non-perturbative properties, such as chiral symmetry breaking and confinement in four-dimensional (4D) space. It is reasonable to also investigate these form factors and polarization moments in the framework of the hard-wall model. Therefore, in this study, we consider the profile function, EM form factors, and tensor polarization observables of the deuteron within the hard-wall model of AdS/QCD. In the hard-wall model, the confinement and chiral symmetry breaking properties of QCD and the finiteness condition of the 5D action are provided by cutoff AdS space at the IR boundary \cite{35,36,37,38,39,40,41}.
	
	In Section II, we present the basic elements of the hard-wall model of the deuteron. We introduce a vector field with twist $\tau=6$ describing the deuteron, derive an equation of motion (EOM) for this field, and find a profile function by solving it. We also fix the IR boundary value according to the ground state mass of the deuteron. In Section III, we introduce an effective action for the bulk field interactions, derive matrix elements for the deuteron EM current, and find the charge monopole, quadrupole, and magnetic dipole form factors and tensor polarization observables of the deuteron. In Section IV, we plot the momentum square $Q^2$ dependence of these form factors and tensor polarization observables and compare our results with the experimental data and soft-wall model results. In Section V, we summarize our results on deuteron EM form factors and tensor polarization observables.

	\section{Hard-wall model of AdS/QCD }
	The finiteness condition of 5D action in AdS space is provided by cutoff at the IR boundary in the hard-wall model. The confinement and chiral symmetry breaking properties of QCD are ensured by this cut-off. The action for the hard-wall model is expressed as \cite{35,36,37,38,39,40,41}:
	\begin{equation}
	S=\int_{0}\limits ^{z_m} d^4x dz\sqrt{G} \mathcal{L}\left(x,z\right),
	\label{1}
	\end{equation}
	where $G=|\det g_{MN}|$ is the determinant of the $g_{MN}$ metric of the AdS space $(M,N=0,1,2,3,5)$, $\mathcal{L}\left(x,z\right)$ is the Lagrangian, and the $z$ coordinate varies in the range $\epsilon\leq z\leq z_{m}\left(\epsilon\rightarrow0\right)$, where $\epsilon\rightarrow0$ is ultraviolet (UV), and $z_{m}$ is the IR boundary of AdS space. The $AdS_{5}$ metric in Poincare coordinates is written as
	\begin{equation}
	ds^{2}=\frac{R^{2}}{z^{2}}\left(-dz^{2}+\eta_{\mu \nu}dx^{\mu}\mu dx^{\nu}\right)=g_{MN}dx^{M}dx^{N}, \nonumber
    \end{equation}
    \begin{equation}
	\quad \mu,\nu=0,1,2,3, 
	\label{2}
	\end{equation}

	where $\eta_{\mu\nu}=diag(1,-1,-1,-1)$ is a 4D Minkowski metric.
	
	\subsection{Vector field bulk-to-boundary propagator}
	
	The bulk-to-boundary propagator of the vector field in the framework of the hard-wall AdS/QCD model is given in Refs. \cite{35,36,37,38,39}. A vector field in the bulk is introduced as a sum of the gauge fields $A_L^M$ and $A_{R}^{M}$, which transform as left and right chiral fields under the $SU(2)_L\times SU(2)_R$ chiral symmetry group of the model \cite{35,36,37,38,39}: $V^M=\frac{1}{\sqrt{2}}\left(A_L^M +A_{R}^{M}\right)$. The UV boundary value of the vector field corresponds to the photon wave function. The 5D bulk action for the free vector field sector is written as \cite{35,36,37,38,39}:
	\begin{equation}
	S=\int\limits_{0}^{z_m}d^5x \sqrt{G} \left(-\frac{1}{2g_5^2}\right) Tr\left(F_{MN}F^{MN}\right),
	\label{3}
	\end{equation}
	where $F_{MN}=\partial_MV_N-\partial_NV_M$ is a field stress tensor of an EM field, $V_M=V_M^at^a$, $t^a=\sigma^a/2$, and $\sigma^a$ are Pauli matrices, $g_5^2=12\pi^2/N_c$ is the 5D coupling constant, and for the dual boundary $SU(2)$ gauge group $g_5=2\pi$, $N_c$ is the number of colors \cite{35,36,37,38,39}.
	
	For the $V_z=0$ gauge choice, the EOM obtained from the action in Eq. (\ref{3}) for the Fourier components $\widetilde{V}_{\mu}^a(q,z)$ is  \cite{35,36,37,38,39}:
	\begin{equation}
	z\partial_z\left(\frac{1}{z}\partial_z\widetilde{V}_{\mu}^a(q,z)\right)+q^{2}\widetilde{V}_{\mu}^a(q,z)=0.
	\label{4}
	\end{equation}
	For the $\widetilde{V}_{\mu}^a(q,z)=V_{\mu}^a(q)V(q,z)$ separation, the $V(q,0)=1$ UV and $V'(q,z_{0})=0$ IR boundary conditions are imposed on the solution of Eq. (\ref{4}) for the vector field. By solving the EOM in Eq. (\ref{4}), the bulk-to-boundary propagator of the vector field is found as \cite{35,37,38,39} :
	\begin{equation}
	V\left(q,z\right)=\frac{\pi}{2}qz\left[\frac{Y_{0}\left( qz_{m}\right)}{J_{0}\left(qz_{m}\right) }J_{1}\left(qz\right)-Y_{1}\left(qz\right)
	\right],
	\label{5}
	\end{equation}
	where $Y_{0}\left(qz_{m}\right)$ and $Y_{1}\left(qz\right)$ are second kind Bessel functions, $J_{0}\left(qz_{m}\right)$ and $J_{1}\left(qz\right)$ are first kind Bessel functions, and $q$ is the 4D momentum. The solution (\ref{5}) in the $q^{2}=-Q^{2}<0$ domain acquires the form~\cite{35}:
	\begin{equation}
	V\left(Q,z\right)=\frac{\pi}{2}Qz\left(\frac{K_{0}\left(Qz_{m}\right)}{I_{0}\left(Qz_{m}\right)}I_{1}\left(Qz\right)+K_{1}\left(Qz\right)\right),
	\label{6}
	\end{equation}
	where $K_{0}\left(Qz_{m}\right)$ and $K_{1}\left(Qz\right)$ are second kind Bessel functions, and $I_{0}\left(Qz_{m}\right)$ and $I_{1}\left(Qz\right)$ are first kind modified Bessel functions.
	
	\subsection{Deuteron profile function}
	In this subsection, we find the deuteron profile function within the hard-wall AdS/QCD model. Similar to the deuteron description in the soft-wall model \cite{19,20}, we express an action for another bulk vector field $F$ with the twist $\tau=6$, which has a UV boundary value corresponding to the deuteron wavefunction:
	\begin{equation}
	S=\int\limits _{0}^{z_{m}}d^4 x dz \sqrt{G}\left[-(D^M F_N)^{+} \left(x,z\right) D_{M} F^{N}\left(x,z\right)+\mu^{2} F_M^+\left(x,z\right)F^{M}\left(x,z\right)\right],
	\label{7}
	\end{equation}
	where $D^{M}=\partial^{M}-ieV^{M}\left(x,z\right)$ is a covariant derivative, $V^{M}\left(x,z\right)$ is vector field dual to the EM field,
	\begin{equation}
	\mu^{2}=\frac{\left(\Delta-1\right)\left(\Delta-3\right)}{R^{2}}=\frac{\left(\tau+L-1\right)\left(\tau+L-3\right)}{R^{2}}=\frac{\left(L+5\right)\left(L+3\right)}{R^{2}} \nonumber
	\end{equation}
	is the 5D mass, $\Delta=\tau+L$ is the conformal dimension of the $F^{N} \left(x,z\right)$ field, and $L$ is the orbital angular momentum.
	
	We use the axial gauge condition for the $F$ field $F^{z}\left(x,z\right)=0$. For the free deuteron field, the Lagrange-Euler EOM is
	\begin{equation}
	\partial_R\left(\frac{\partial{L}}{\partial\left(\partial_{R}F_B^{+}\right)}\right)-\frac{\partial{L}}{\partial{F_{B}^{+}}}=0,
	\label{8}
	\end{equation}
	\begin{equation}
	\left[-\frac{\partial^{2}}{\partial z^{2}}+\frac{1}{z} \frac{\partial }{\partial z}+\frac{\left(L+4\right)^2-1}{z^{2}}\right] F^{\nu}\left(x,z\right)=M^{2}_{D,n} F^{\nu}\left(x,z\right).
	\label{9}
	\end{equation}
	Note that in Eq. (\ref{9}) the $\partial_{\alpha} \partial_{\beta} F^{\nu}\left(x,z\right) \eta^{\alpha\beta}=-M^{2}_{D,n} F^{\nu}\left(x,z\right)$ 4D EOM is applied.
	
     Kaluza-Klein (KK) decomposition of the $F$ deuteron field
	\begin{equation}
	F^{\nu}\left(x,z\right)=\exp^{-\frac{A\left(z\right)}{2}}\sum_{n}F_{n}^{\nu}\left(x\right) \Phi_{n}\left(z\right)
	\label{10}
	\end{equation}
	is used along with the $F_{n}^{\nu}\left(x\right)$ tower of KK components dual to deuteron states with the radial quantum number $n$.
	
	Thus, we obtain a Schrodinger-type EOM for the $\Phi_{n}\left(z\right)$ bulk profile.
	\begin{equation}
	\left[-\frac{d^{2}}{dz^{2}}+\frac{4\left(L+4\right)^{2}-1}{4z^{2}}\right]\Phi_{n}\left(z\right)=M^{2}_{D,n} \Phi_{n}\left(z\right),
	\label{11}
	\end{equation}
	where $M_n$ is the mass spectrum of the deuteron. Solving this equation for $L=1$, we find the profile function of the $n$-th mode $\Phi_{n}\left(z\right)$ deuteron.
	\begin{equation}
	\Phi_{n}\left(z\right)=b_{1}\sqrt{z} J_{5}\left(M_{D,n} z\right)+b_{2}\sqrt{z} Y_{5}\left(M_{D,n} z\right),
	\end{equation}
	where $J_{5}\left(M_{D,n} z\right)$ and $Y_{5}\left(M_{D,n} z\right)$ are the Bessel functions of the first and second kinds, respectively. The UV boundary condition requests $b_{2}$ to be $b_{2}=0$, and the normalization condition gives us the $b_{1}$ constant as  
	\begin{equation}
	b_{1}=\frac{1}{\sqrt{\int_{0}^{z_{m}}dz\left[J_{5}\left(M_Dz\right)\right]^{2}}}.  \label{13}
	\end{equation}
	Thus, the profile function of the deuteron in the framework of the hard-wall AdS/QCD model accepts the following form: 
	\begin{equation} 
	\Phi\left(z\right)=\frac{\sqrt{z}J_{5}\left(M_Dz\right)}{\sqrt{\int_{0}^{z_{m}}dz\left[J_{5}\left(M_Dz\right)\right]^{2}}}.
	\label{14}
	\end{equation}
	The IR boundary condition leads to $J_{5}\left(M_D z_m\right)=J_{5}\left(\alpha_{0}\right)=0$, where $M_D$ is the ground state deuteron mass ($M_{D}=1,876$ $GeV=9.508$ $fm^{-1}$) and $\alpha_{0}$ is the first zero of $J_5$ $(\alpha_{0}=8.77)$. We fix the value of the IR boundary at $z_m=\frac{\alpha_{0}}{M_D}=4.6756$ $GeV^{-1}=0.9225$ $fm$.

	\section{Deuteron EM form-factors and tensor polarization observables within the hard-wall AdS/QCD model}
	
	\subsection{Effective action for the bulk interaction}
	Here, we derive the deuteron EM form factors and tensor polarization observables in the framework of the hard-wall model. Hence, we present an effective action, which includes all interactions between the EM and deuteron fields in the bulk of  AdS space. For clarity, we separately present all bulk Lagrangian terms used in the Refs. \cite{19,20} that contribute to the invariant form factors $G_{1}\left( Q^{2}\right)$, $G_{2}\left( Q^{2}\right)$, and $G_{3}\left( Q^{2}\right)$:
	
	1) A minimal bulk gauge action term $\textit{S}^{\left(1\right)}$, which describes the minimal gauge interaction of the EM field with the deuteron,
	\begin{equation}
	\textit{S}^{\left(1\right)}=\int d^{4}x\ \int\limits_{0}^{z_{M}}dz \sqrt{G}\left[-D^M F_N^+ \left(x,z\right) D_{M} F^{N}\left(x,z\right)\right],
	\label{15}
	\end{equation}
	2) the action term $\textit{S}^{\left(2\right)}$, which leads to the magnetic dipole form factor of the deuteron,
	\begin{equation}
	\textit{S}^{\left(2\right)}=\int d^{4}x\ \int_{0}^{z_{M}}dz \sqrt{G}\left[-ic_{2}F^{MN}\left(x,z\right) F^{+}_{M}\left(x,z\right)F_N\left(x,z\right)\right],
	\label{16}
	\end{equation}
	3) a non-minimal bulk gauge action term $\textit{S}^{\left(3\right)}$, which was introduced in \cite{19,20} and contributes to the charge monopole and quadrupole form factors of the deuteron,
	\begin{eqnarray}
	\textit{S}^{\left(3\right)}=\int d^{4}x\ \int_{0}^{z_{M}} dz
	\sqrt{G}\frac{c_{3}}{4M^2_D}\exp^{2A\left(z\right)}\partial^MF^{NK}\left(x,z\right)\left[i\partial_{K}F^{+}_{M}\left(x,z\right)F_N\left(x,z\right)-\nonumber \right.\\
	-F^{+}_{M}\left(x,z\right)i\partial_K F_N\left(x,z\right)+H.C.\left.\right]. \label{17}
	\end{eqnarray}
	The general bulk action for the interaction will be a sum of these terms.
	\begin{equation}
	\textit{S}_{int}=\textit{S}^{\left(1\right)}+\textit{S}^{\left(2\right)}+\textit{S}^{\left(3\right)}.
	\label{18}
	\end{equation}
	After performing several calculations, the action terms accept the following form:
	\begin{equation}
	\textit{S}^{\left(1\right)}=\int d^{4}x\ \int_{0}^{z_{M}}dz \sqrt{G} e V_{\mu}\left(x,z\right)\left[i\partial^{\mu} F_{\nu}^+ \left(x,z\right) F^{\nu}\left(x,z\right)-F_{\nu}^+ \left(x,z\right) i\partial^{\mu}  F^{\nu}\left(x,z\right)\right],
	\label{19}
	\end{equation}
	\begin{eqnarray}
	\textit{S}^{\left(2\right)}=\int d^{4}x\ \int_{0}^{z_{M}}dz\sqrt{G}c_{2}\left[i\partial_{\nu}\left(x,z\right)V_{\mu}\left(x,z\right) F^{\mu+}\left(x,z\right)F^{N}\left(x,z\right)-\nonumber \right. \\
	-i\partial_{\mu}\left(x,z\right)V_{\nu}\left(x,z\right)F^{\mu+}\left(x,z\right)F^{N}\left(x,z\right) \left.\right], \label{20}
	\end{eqnarray}
	\begin{eqnarray}
	\textit{S}^{\left(3\right)}=\int d^{4}x\ \int_{0}^{z_{M}}dz\sqrt{G}\frac{c_{3}}{2M^{2}_{D}}\exp^{2A\left(z\right)}\left[-2\partial^{\mu}\partial^{\nu}V^{K}\left(x,z\right)eV_{k}F_{\mu}^{+}\left(x,z\right)F_{\nu}\left(x,z\right)- \nonumber \right.  \\
	-\partial^{\mu}\partial^{\nu}V^{K}\left(x,z\right)F_{\mu}^{+}\left(x,z\right)i\partial_{k}F_{\nu}\left(x,z\right)+2\partial^{\mu}\partial^{k}V^{\nu}\left(x,z\right)F_{\mu}^{+}\left(x,z\right)eV_{k}\left(x,z\right)F_{\nu}\left(x,z\right)+ \nonumber \\
	+\partial^{\mu}\partial^{\nu}V^{K}\left(x,z\right)i\partial_{k}F_{\mu}^{+}\left(x,z\right)F_{\nu}\left(x,z\right)-\partial^{\mu}\partial^{k}V^{\nu}\left(x,z\right)i\partial_{k}F_{\mu}^{+}\left(x,z\right)F_{\nu}\left(x,z\right)+ \nonumber \\
	+\partial^{\mu}\partial^{k}V^{\nu}\left(x,z\right)F_{\mu}^{+}\left(x,z\right)i\partial_{k}F_{\nu}\left(x,z\right)-2\partial^{\mu}\partial^{k}V^{\nu}\left(x,z\right)F_{\mu}\left(x,z\right)eV_{k}\left(x,z\right)F_{\nu}^{+}\left(x,z\right)+ \nonumber \\
	+2\partial^{\mu}\partial^{\nu}V^{k}\left(x,z\right)eV_{k}\left(x,z\right)F_{\mu}\left(x,z\right)F_{\nu}^{+}\left(x,z\right)-\partial^{\mu}\partial^{\nu}V^{k}\left(x,z\right)F_{\mu}\left(x,z\right)i\partial_{k}F_{\nu}^{+}\left(x,z\right)+ \nonumber \\
	+\partial^{\mu}\partial^{k}V^{\nu}\left(x,z\right)F_{\mu}\left(x,z\right)i\partial_{k}F_{\nu}^{+}\left(x,z\right)+\partial^{\mu}\partial^{\nu}V^{k}\left(x,z\right)i\partial_{k}F_{\mu}\left(x,z\right)F_{\nu}^{+}\left(x,z\right)- \nonumber  \\
	-\partial^{\mu}\partial^{k}V^{\nu}\left(x,z\right)i\partial_{k}F_{\mu}\left(x,z\right)F_{\nu}^{+}\left(x,z\right)\left. \right]. \label{21}
	\end{eqnarray}
	We perform the decomposition (\ref{10}) in Eqs. (\ref{19})-(\ref{21}) and apply a Fourier transformation to the vector $V\left(x,z\right)$ and deuteron $F_{\nu}\left(x\right)$, $F_{\nu}^{+}\left(x\right)$ fields.
	\begin{equation}
	V_{\mu}\left(x,z\right)=\int \frac{d^{4}q}{\left(2\pi\right)^4}\exp^{-iqx}V_{\mu}\left(q\right)V\left(q,z\right)
	\label{22}.
	\end{equation}
	\begin{equation}
	F_{\nu}\left(x\right)=\int \frac{d^{4}p}{\left(2\pi\right)^{4}}e^{-ipx} \epsilon_{\nu}\left(p\right),
	\label{23}
	\end{equation}
	\begin{equation}
	F_{\nu}^{+}\left(x\right)=\int \frac{d^{4}p^{'}}{\left(2\pi\right)^{4}}e^{ip^{'}x}\epsilon_{\nu}^{+}\left(p^{'}\right),
	\label{24}
	\end{equation}
	where $p$ and $p^{'}$ are the four-momenta, and $\epsilon$ and $\epsilon^{+}$ are the polarizations of the initial and final deuterons, respectively. Substituting Eqs. (\ref{10}) and (\ref{22})-(\ref{24}) into Eqs.(\ref{19})-(\ref{21}), we  obtain action terms for the bulk deuteron-EM field interactions.
	\begin{eqnarray}
	\textit{S}^{\left(1\right)}=-\left(2\pi\right)^{4}\int\frac{d^{4}p}{\left(2\pi\right)^{4}}\int\frac{d^{4}p^{'}}{\left(2\pi\right)^{4}}\int\frac{d^{4}q}{\left(2\pi\right)^{4}}\delta^{4}\left(p+q-p{'}\right)eV_{\mu}\left(q\right)\int dz V\left(q,z\right)\Phi^{2}\left(z\right) \times \nonumber\\
	\times\epsilon^{+}(p^{'})\epsilon\left(p\right)(p+p^{'})^{\mu}), \label{25}
	\end{eqnarray}
	\begin{eqnarray}
	\textit{S}^{\left(2\right)}=-c_{2}\left(2\pi\right)^{4}\int\frac{d^{4}p}{\left(2\pi\right)^{4}}\int\frac{d^{4}p^{'}}{\left(2\pi\right)^{4}}\int\frac{d^{4}q}{\left(2\pi\right)^{4}}\delta^{4}(p+q-p{'}) V_{\mu}\left(q\right)\int dz V\left(q,z\right) \Phi^{2}\left(z\right)\times  \nonumber\\
	\times(\epsilon^{\mu}\left(p\right)\epsilon^{+}(p^{'}) q-\epsilon^{+\mu}(p^{'})\epsilon\left(p\right)\cdot q)(p,p^{'}), \label{26}
	\end{eqnarray}
	\begin{eqnarray}
	\textit{S}^{\left(3\right)}=\left(2\pi\right)^{4} \frac{c_{3}}{2M^{2}} \int\frac{d^{4}p} {\left(2\pi\right)^{4}}\int\frac{d^{4}p^{'}}{\left(2\pi\right)^{4}}\int\frac{d^{4}q}{\left(2\pi\right)^{4}}\delta^{4}(p+q-p{'}) V_{\mu}\left(q\right)  \int dz V\left(q,z\right) \Phi^{2}\left(z\right)\times  \nonumber\\
	\times\epsilon^{+}(p^{'})\cdot q\epsilon\left(p\right)\cdot q(p+p^{'})^{\mu}. \label{27}
	\end{eqnarray}
	Action terms (\ref{25})-(\ref{27}) contribute to the $G_{1}\left( Q^{2}\right)$, $G_{2}\left( Q^{2}\right)$, and $G_{3}\left( Q^{2}\right)$ form factors of the deuteron.
	
	\subsection{EM current, matrix element, and EM form-factors}
	
	According to AdS/CFT correspondence, the deuteron EM current can be found by taking a variation from the generating functional $Z=e^{iS_{int}}$ over the vector field $V_{\mu}\left(q\right)$ \cite{38}.
	\begin{equation}
	\left\langle J^{\mu}\left(p,p^{'}\right)\right\rangle=-\frac{\delta e^{iS_{int}}}{\delta V_{\mu}\left(q\right)}|_{V_{\mu}=0}.
	\label{28}
	\end{equation}
	According to Eq. (\ref{28}), the action terms of Eqs. (25)-(27) give the corresponding current terms,
	\begin{eqnarray}
	J^{\mu\left(1\right)}(p,p^{'})=-\int dz V\left(Q,z\right) \phi^{2}\left(z\right) \epsilon^{+}(p^{'}) \cdot \epsilon\left(p\right)(p+p^{'})^{\mu}=\nonumber \\
	=-G_{1}\left(Q^{2}\right)\epsilon^{+}(p^{'}) \cdot \epsilon\left(p\right)(p+p^{'})^{\mu},\label{29}
	\end{eqnarray}
	\begin{eqnarray}
	J^{\mu(2)}(p,p^{'})=-c_{2} \int dz V\left(Q,z\right) \phi^{2}\left(z\right) (\epsilon^{\mu}(p) \epsilon^{+}(p^{'})\cdot q-\epsilon^{+\mu}(p^{'}) \epsilon\left(p\right)\cdot q)=\nonumber \\
	=-G_{2}\left(Q^{2}\right)(\epsilon^{\mu}\left(p\right) \epsilon^{+}(p^{'})\cdot q-\epsilon^{+\mu}(p^{'}) \epsilon\left(p\right)\cdot q), \label{30}
	\end{eqnarray}
	\begin{eqnarray}
	J^{\mu\left(3\right)}(p,p^{'})=\frac{c_{3}}{2M^{2}_{D}}\int dz V\left(Q,z\right) \phi^{2}\left(z\right) \epsilon^{+}(p^{'})\cdot q \epsilon\left(p\right)\cdot q (p+p^{'})^{\mu}=\nonumber \\
	=\frac{G_{3}\left(Q^{2}\right)}{2M^{2}_{D}}\epsilon^{+}(p^{'})\cdot q \epsilon\left(p\right)\cdot q (p+p^{'})^{\mu}. \label{31}
	\end{eqnarray}
	The current conservation and $P$- and $C$-invariances provide three EM form factors of the deuteron, known as the charge monopole $G_C(Q^2)$, quadrupole $G_Q(Q^2)$ and magnetic dipole $G_M(Q^2)$ form factors.
	
	The EM current of a $e+d\rightarrow e+d$ electron-deuteron elastic scattering process is written in terms of $G_{i}\left(Q^{2}\right)$ form factors:
	\begin{eqnarray}
	J^{\mu}(p,p^{'})=-\left(G_{1}\left(Q^{2}\right)\epsilon^{+}(p^{'})\cdot \epsilon\left(p\right)-\frac{G_{3}\left(Q^{2}\right)}{2M^2_{D}}\epsilon^{+}(p^{'})\cdot q  \epsilon\left(p\right)\cdot q \right)(p+p^{'})^{\mu}-\nonumber \\
	-G_{2}\left(Q^{2}\right)\left(\epsilon^{\mu}\left(p\right) \epsilon^{+}(p^{'})\cdot q-\epsilon^{+\mu}(p^{'}) \epsilon\left(p\right)\cdot q\right).\label{32}
	\end{eqnarray}
	
	According to AdS/CFT correspondence, the sum of current terms in Eqs. (29)-(31) is identified in the current form (32). 
	
	$G_{1}\left(Q^{2}\right)$, $G_{2}\left(Q^{2}\right)$ and $G_{3}\left(Q^{2}\right)$ are deuteron form factors that depend only on the virtual photon four-momentum. Moreover, assuming hermiticity, they are real. The explicit expressions of these form factors are found by comparing Eqs. (\ref{29})-(\ref{31}) with Eq. (\ref{32}) :
	\begin{eqnarray}
	G_1(Q^2)=\int dz V\left(Q,z\right) \Phi^{2}\left(z\right), \nonumber \\ 
	G_2(Q^2)=c_{2} \int dz V\left(Q,z\right) \Phi^{2}\left(z\right), \nonumber \\ 
	G_3(Q^2)=c_{3} \int dz V\left(Q,z\right) \Phi^{2}\left(z\right).
	\label{33}
	\end{eqnarray}
	
	The charge monopole $G_C(Q^2)$, quadrupole $G_Q(Q^2)$, and magnetic $G_M(Q^2)$ form factors are related to the $G_1(Q^2)$, $G_2(Q^2)$, and $G_3(Q^2)$ form-factors as ~\cite{3,17,19,20}:
	\begin{eqnarray}
	G_C(Q^2)=G_1(Q^2)+\frac{2}{3}\eta_{d}G_Q(Q^2), \nonumber \\ 
	G_Q(Q^2)=G_1(Q^2)-G_2(Q^2)+\left(1+\eta_{d}\right)G_3(Q^2), \nonumber \\
	G_M(Q^2)=G_2(Q^2), \label{34}
	\end{eqnarray}
	where $\eta_{d}=\frac{Q^2}{4M^2_D}$.
	At $Q^2=0$, the charge monopole $G_C(Q^2)$, quadrupole $G_Q(Q^2)$, and magnetic dipole $G_M(Q^2)$ form factors are normalized experimentally as follows \cite{17}:
	\begin{eqnarray}
	G_C(0)=1,  \nonumber \\ 
	G_Q(0)=M^{2}_{D}Q_{D}=25.83, \nonumber \\ 
	G_M(0)=\frac{M_{D}}{m_{N}}\mu_D=1.714,   \label{35}
	\end{eqnarray}
	where $m_{N}$ is the nucleon mass, $Q_{D}$ is a quadrupole, and $\mu_D$ is the magnetic moment of the deuteron.
	Using relation (\ref{34}) and the normalization of the $G_Q(Q^2)$ and $G_M(Q^2)$ form-factors at $Q^2=0$ in Eq. (\ref{35}) , we fix the value of the constants $c_2$ and $c_3$ as $c_2=1.67066$ and $c_3=22.507$, respectively. 
	
	\subsection{Tensor polarization observables of the deuteron}

	In elastic $e+d\rightarrow e+d$ electron-deuteron scattering, two unpolarized elastic structure functions, $A(Q^2)$ and $B(Q^2)$, are applied in the description of the scattering process. The two unpolarized elastic structure functions are defined using the $G_C(Q^2)$, $G_Q(Q^2)$, and $G_M(Q^2)$ form factors as follows  \cite{3,17,19,21,22,24,29}:
	\begin{eqnarray}
	A(Q^2)=G^2_C(Q^2)+\frac{2}{3}\eta_{d}G^2_M(Q^2)+\frac{8}{9}\eta^2_{d}G^2_Q(Q^2), \nonumber \\ 
	B(Q^2)=\frac{4}{3}\eta_{d}(1+\eta_{d})G^2_M(Q^2). 
	\label{36}
	\end{eqnarray}
	The $B(Q^2)$ unpolarized elastic structure function depends only on the $G^2_M(Q^2)$ form factor. Therefore, this form factor can be determined by a Rosenbluth separation of $A(Q^2)$ and $B(Q^2)$ \cite{3,17}: 
	\begin{equation}
	\frac{d\sigma}{d\Omega}=\left(\frac{d\sigma}{d\Omega}\right)_{Mott} S\left(Q^2\right),
	\label {37}
	\end{equation}
	or by a cross section measurement at $\theta = 180^{o}$, where $\left(\frac{d\sigma}{d\Omega}\right)_{Mott}$ is the Mott cross section, $\theta$ is the electron scattering angle, and $S=A(Q^2)+B(Q^2)$~\cite{3,17}. Because the first unpolarized elastic structure function $A(Q^2)$ depends on all three form factors, only a quadratic combination of $G_C(Q^2)$ and $G_Q(Q^2)$ can be determined from the unpolarized cross section. A complete separation of the form factors requires the measurement of at least one tensor polarization observation, such as $T_{20}$, $T_{21}$, and $T_{22}$, and all of the polarization observables depend on the momentum square $Q^2$ and scattering angle $\theta$ \cite{3,17,19,21,24,29,30,31}:
	\begin{eqnarray}
	T_{20}(Q^2,\theta)=-\frac{1}{\sqrt{2}S(Q^2)}\left[\frac{8}{3}\eta_{d}(Q^2)G_C(Q^2)G_Q(Q^2)+\frac{8}{9}\eta_{d}^{2}(Q^2)G_Q^2(Q^2)+ \right. \nonumber \\
	+\left.\frac{1}{3}\eta_{d}(Q^2)\left(1+2\left(1+\eta_{d}(Q^2)\right)\tan^{2}\frac{\theta}{2}\right)G_M(Q^2)\right],   \nonumber \\ 
	T_{21}(Q^2,\theta)=-\frac{2}{\sqrt{3}S(Q^2)}\eta_{d}(Q^2)\sqrt{\eta_{d}(Q^2)\left(1+\eta_{d}(Q^2)\sin^{2}\frac{\theta}{2}\right)}\frac{G_M(Q^2)G_Q(Q^2)}{\cos\frac{\theta}{2} }, \nonumber \\
	T_{22}(Q^2,\theta)=-\frac{1}{2\sqrt{3}S(Q^2)}\eta_{d}(Q^2)G_M^2(Q^2) 
	\label{38} 
	\end{eqnarray} 
	Because the first $T_{20}$ measurement \cite{32} was performed close to $\theta= 70^{o}$, observables are quoted at this angle. For experiments not performed at $70^{o}$, the $T_{20}$ observable is extrapolated to this angle using the structure functions. It is useful to use an alternative quantity, $\tilde{T}_{20}(Q^2)$, which is independent of scattering angle $\theta$ and thus depends only on $Q^{2}$~\cite{17,19,31}:
	\begin{eqnarray}
	\tilde{T}_{20}(Q^2)=-\frac{\eta_{d}(Q^2)}{\sqrt{2}}\frac{3\beta(Q^2)+\eta_{d}(Q^2)}{\frac{9}{8}\beta^{2}(Q^2)+\eta_{d}^{2}(Q^2)}, \nonumber \\ 
	\tilde{T}_{20R}(Q^2)=\frac{G_{Q}(Q^2)}{G_{Q}(0)}\frac{G_{C}(Q^2)+\frac{\eta_{d}}{3}G_{Q}(Q^2)}{G_{C}^{2}(Q^2)+\frac{8}{9}\eta_{d}^{2}G_{Q}^2(Q^2)}, \label{39} 
	\end{eqnarray}
	where $\beta(Q^2)=\frac{G_{C}(Q^2)}{G_{M}(Q^2)}$ is the ratio of the charge monopole and magnetic dipole EM form factors of a deuteron, and $\tilde{T}_{20R}(Q^2)=1$, normalized at $Q^2=0$.
	
	\section{Numerical results for the deuteron form factors}
	We numerically calculate integrals for $G_C\left(Q^2\right)$, $G_Q\left(Q^2\right)$, and $G_M\left(Q^2\right)$ and present their $Q$ dependence for the EM form factors of the deuteron in Fig. 1. We also numerically calculate integrals for the $T_{20}\left(Q^2\right)$, $T_{20R}\left(Q^2\right)$, $T_{21}\left(Q^2\right)$, and $T_{22}\left(Q^2\right)$ tensor polarization observables and present their $Q$ dependence for these observables in Fig. 2. To perform numerical integration, we fix the IR boundary value $z_m$ within the hard-wall AdS/QCD model from the zeros of the first kind Bessel function $J_5$ as $z_m= 0.9225$ $fm$. The value $\theta=70^{o}$ is taken from experiments \cite{33,34}, and the values of the nucleon and deuteron masses are $m_{N}=4.764$ $fm^{-1}$ and $M_{D}=9.508$ $fm^{-1}$, respectively. 
	\begin{figure}
		\includegraphics[scale=0.30]{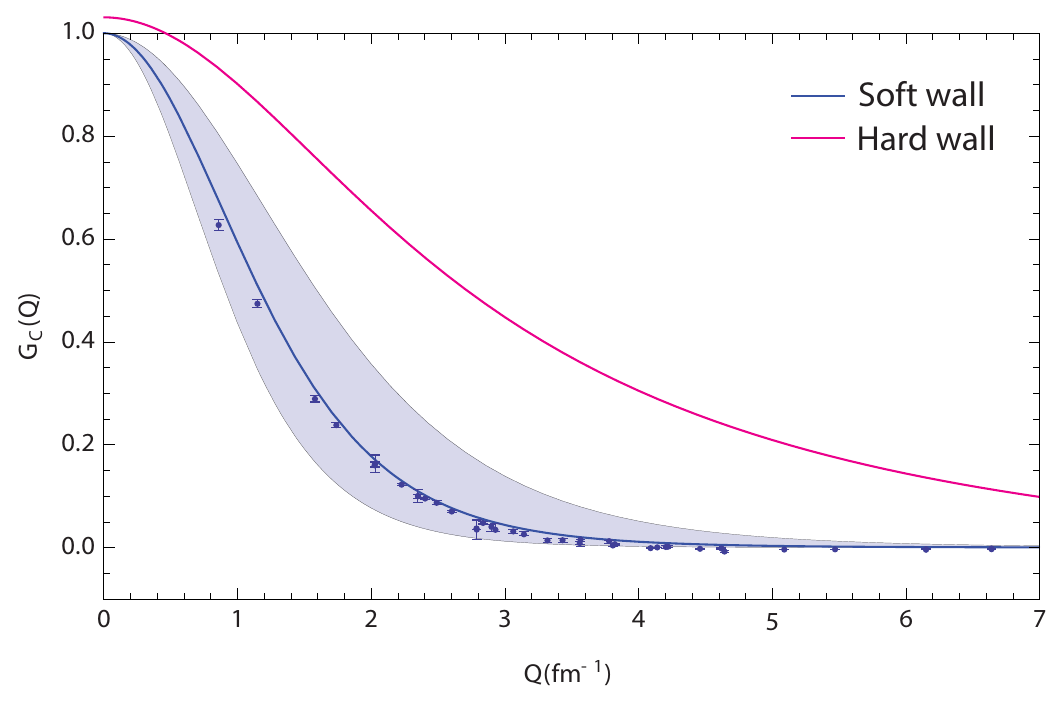}
		\includegraphics[scale=0.30]{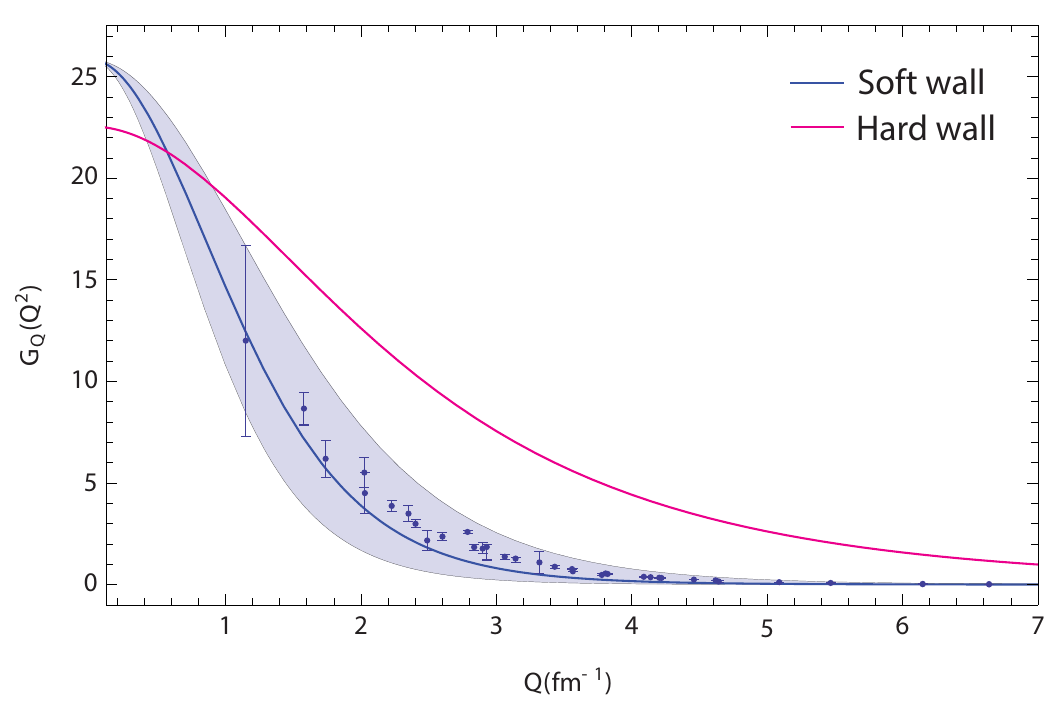}
		\includegraphics[scale=0.30]{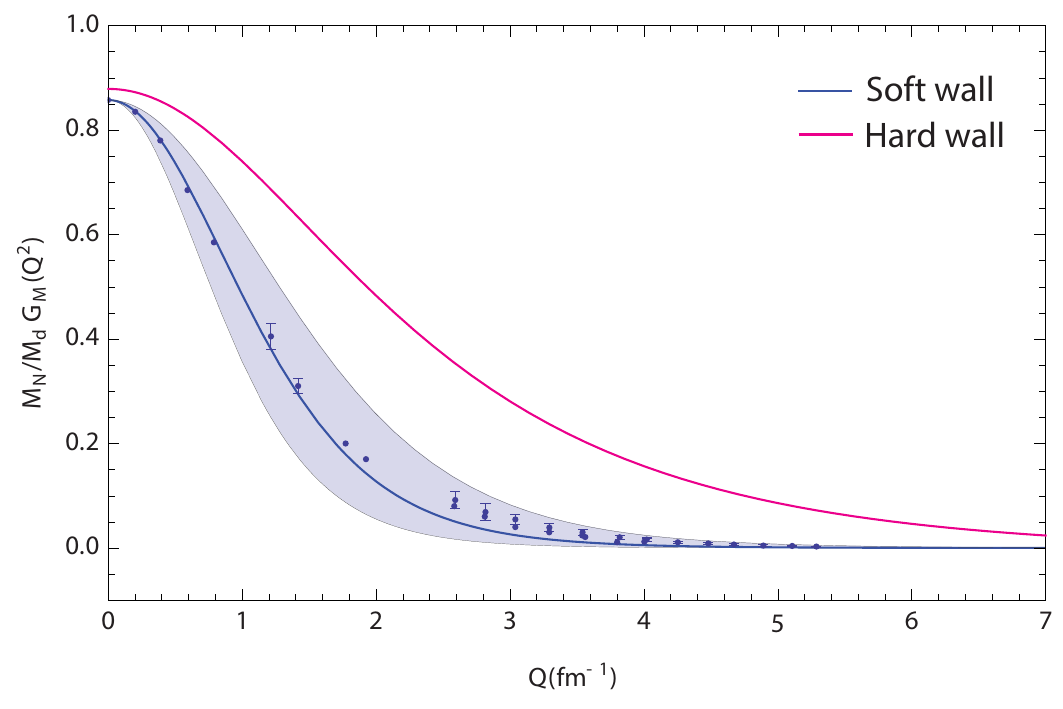}
		\caption {$\left(color online\right)$ $G_C\left(Q^2\right)$ charge monopole, $G_Q\left(Q^2\right)$ quadrupole and $\frac{m_{N}}{M_{D}}G_M\left(Q^2\right)$ magnetic dipole deuteron form factors.}
		\label{fig: 1}
	\end{figure}
	\begin{figure}
		\includegraphics[scale=0.35]{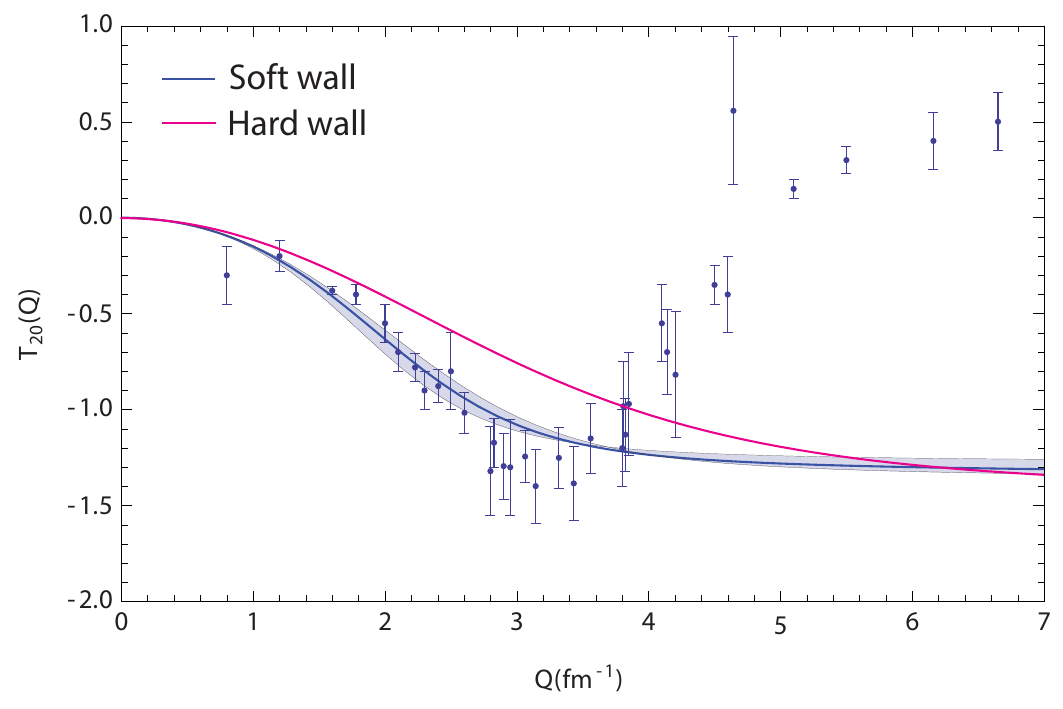}
		\includegraphics[scale=0.35]{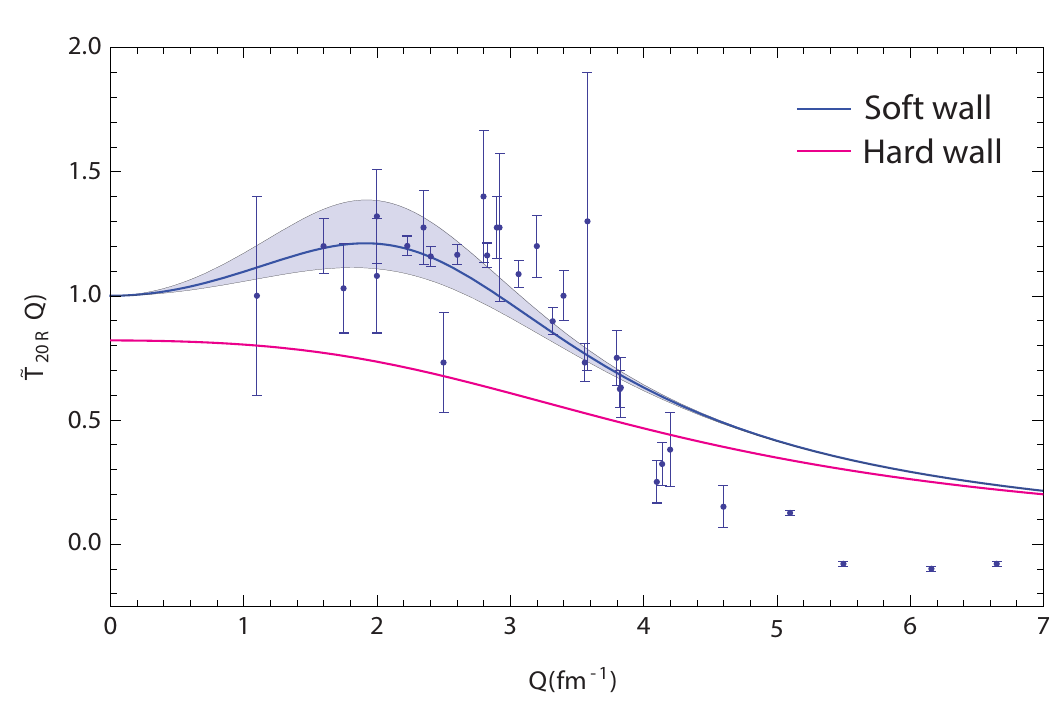}
		\includegraphics[scale=0.35]{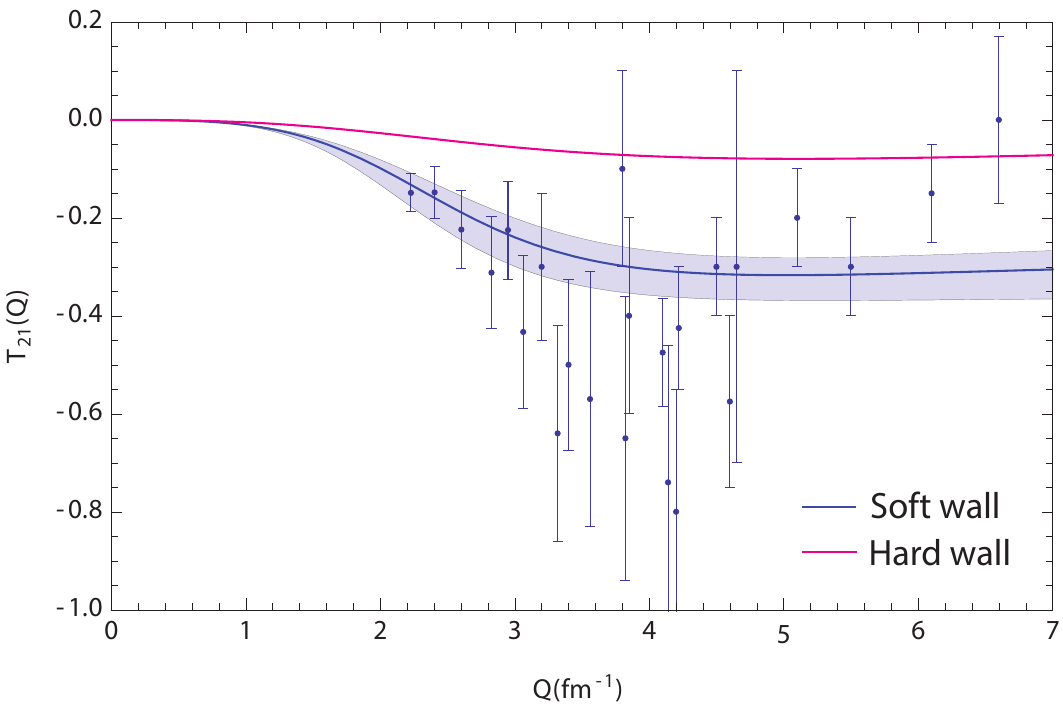}
		\includegraphics[scale=0.35]{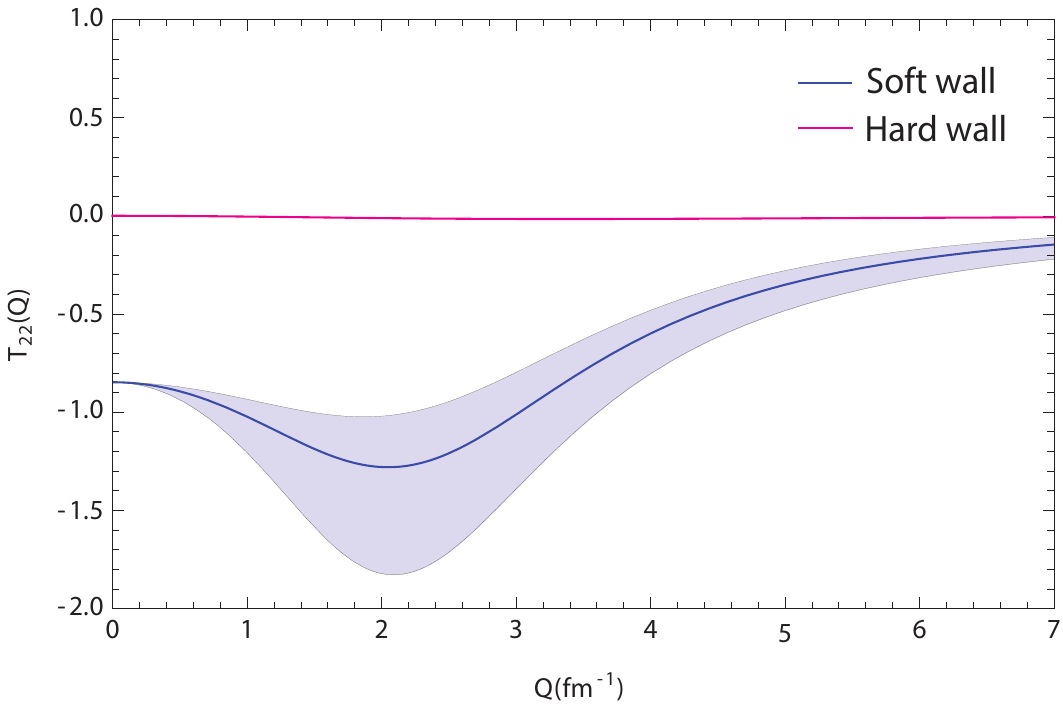}
		\caption {$\left(color online\right)$ $T_{20}\left(Q^2\right)$, $T_{20R}\left(Q^2\right)$, $T_{21}\left(Q^2\right)$ and $T_{22}\left(Q^2\right)$ deuteron tensor-polarized moments.}
		\label{fig: 2}
	\end{figure}
	
	As identified from Fig.1, the hard-wall results for the $G_C\left(Q^2\right)$, $G_Q\left(Q^2\right)$, and $G_M\left(Q^2\right)$ form factor $Q$ dependences exhibit the shape of $1/Q^n$ dependence. Such a dependence is typical of deuteron form factors\cite{1,2,3,4,5,6,7,8,9,10,11,12,13,14,15,16,17,18,19,20,21,22,23,24,25,26,27,28}. Existing experimental data on the $G_C\left(Q^2\right)$, $G_Q\left(Q^2\right)$, and $G_M\left(Q^2\right)$ form factors and tensor polarization observables of the deuteron in Refs. \cite{33,34} are in the $\sim 0.86\leq Q \left(fm^{-1}\right)\leq 6.64$ interval of $Q$. For a detailed analysis, we conduct a comparison with experimental data and soft-wall model results. These data are represented in Fig. 1 and Fig. 2 as dark dots. In graphs, the blue solid lines represent the soft-wall model results obtained in Ref.~\cite{19}, and the red solid lines are our hard-wall model results.

	\section{Summary and Conclusion}
	In this study, we calculate the EM form-factors and tensor polarization observables of the deuteron in the framework of the hard-wall AdS/QCD model and compare the numerical results with those obtained within the soft-wall model. The graphs  of the hard-wall form factors results exhibit similar shapes to those of existing experimental data and the results of the soft-wall model. Regarding the values, the dependencies presented in the form factor graphs exhibit shifts with data and the results. The graphs of $T_{20}$ and $T_{20R}$ are similar to the soft-wall results, whereas the graphs of $T_{21}$ and $T_{22}$ differ from the soft-wall results. As known from studies to data, no holographic model can completely describe all phenomenological quantities. The results of the same quantity often differ significantly between different holographic models \cite{66}. Therefore, probing both the soft and hard-wall models for such problems will be useful.

\end{document}